\documentclass{aa501}
\usepackage{graphicx}
\def\hmpc{~h$^{-1}$ Mpc~}

\def\scsette{SC~1327--312~}
\def\scnove{SC~1329--313~}
\begin{document}
%
%
%-------------------------------------------------------------------------------
%
\title
{\scsette \& \scnove: two galaxy groups in-between a major merging event
observed with Beppo--SAX}
%
%-------------------------------------------------------------------------------
%
\author
{
S.~Bardelli\inst{1} \and
S.~De Grandi\inst{2} \and
S.~Ettori\inst{3} \and
S.~Molendi\inst{4} \and
E.~Zucca\inst{1} \and
S.~Colafrancesco\inst{5} 
}
\institute
{
Osservatorio Astronomico di Bologna, 
via Ranzani 1, I--40127 Bologna, Italy
\and
Osservatorio Astronomico di Brera, Via Bianchi 46, I-23807 Merate
(LC), Italy 
\and
Institute of Astronomy, Madingley Road, CB3 0HA Cambridge, England 
\and
Istituto di Fisica Cosmica ``G.Occhialini'', Via Bassini 15, I-20133
Milano, Italy 
\and
Osservatorio Astronomico di Roma, via Frascati 33, I-00040 
Monteporzio Catone, Italy
}
%
%-------------------------------------------------------------------------------
%
\date{Received 00 - 00 - 0000; accepted 00 - 00 - 0000}
%
%-------------------------------------------------------------------------------
\titlerunning{Groups in a merging structure}
\authorrunning{S. Bardelli et al.}

%-------------------------------------------------------------------------------
\abstract{ 
We present the results of two Beppo-SAX observations of the poor clusters
\scsette and \scnove: these objects are located in a huge structure, formed 
by three ACO clusters, which is probably the remnant of one of the largest 
and best studied major mergings known. Given the fact that these poor 
clusters are in between two interacting ACO clusters, the aim of this work is 
to study the physics of the intracluster medium and to look for the possible 
presence of shocks.
\\
We derived the gas distribution profiles, the global (i.e. within $0.3$ 
\hmpc) temperatures and abundances and the temperature profiles and maps for 
\scsette and \scnove. Also the presence of soft excess in LECS and
hard excess in PDS have been checked.
\\ 
We do not find evidence of regions where the gas is shocked or significantly 
heated. The image of \scsette seems rather symmetric, while the gas profile 
of \scnove shows disturbed, ``comet-like" shaped isophotes, with the tail 
pointing toward A3562 and a compression toward \scsette.
\\
The presence of multiphase gas in \scnove, as claimed by  Hanami et al. 
(\cite{hanami99}) on ASCA data, has been found only at the 2$\sigma$ confidence
level.   
\\
The lack of heating supports the hypothesis that the merging is at a
late stage, after the first core--core encounter, when the main shock front
had the time to travel to the external regions of the main clusters.     
}
%-------------------------------------------------------------------------------
\maketitle
%
%
%-------------------------------------------------------------------------------
%
\section{Introduction}

Cluster mergings are among the most energetic phenomena in the Universe, 
leading to a release of $\sim 10^{50-60}$ erg in a time scale of the order 
of few Gyrs (Sarazin \cite{sarazin00}), but it is not yet clear in detail 
in which way this kinetic energy is dissipated and which is its influence 
on the galaxy population.   
There are some observational features that have been associated with the 
cluster merging phenomenon, like shocks in the hot gas, radio halos, relics 
and wide angle tail radiosources, and the presence of starburst activity
in galaxies, although the precise theoretical description is not yet 
completely assessed. 
\\
The main difficulty in understanding the merging phenomenon is that the
available data allow only studies focussed on single events, mainly in cases 
where the intervening entity has a much smaller mass with respect to the 
main cluster. Moreover, some authors limited the analysis to single 
wavelengths (in particular X-ray band): however, although the X-ray band is 
very powerful for these studies, a multiwavelength analysis would give a 
more global view of the phenomenon.
\\
Another important point is the key role played by the large--scale 
environment. It has been recognized from cosmological simulations that 
rich clusters are accreting groups and other clusters along filaments 
(Colberg et al. \cite{colberg99}), which determine the relative velocity 
and frequency of the impacts.  
Rich superclusters are places where one can find the most spectacular cases 
of such dynamical phenomena: the peculiar velocities due to the enhanced local
density favour cluster-cluster collisions, i.e. involving entities of
similar mass (hereafter ``major mergings"). 
\\
The most remarkable examples of cluster mergings are found in the central
region of the Shapley Concentration, the richest supercluster of clusters 
within 300 \hmpc (Zucca et al. \cite{zucca93}).
At the center of the Shapley Concentration we individuated (Bardelli et al. 
\cite{bardelli94}) two cluster structures (on scales of $\sim 5$ \hmpc) which 
are cluster-cluster mergings at two different stages. 
The most massive structure (dubbed A3558 complex, Bardelli et al. 
\cite{bardelli94}, \cite{bardelli98a}) is probably a merging 
seen just after the first core-core encounter. 
The other complex, dominated by A3528, is formed by two pairs of clusters 
and is probably at the first stages of interaction (Bardelli et al.  
\cite{bardelli01}, Baldi et al. \cite{baldi01}).  
\\
In this paper we present the analysis of Beppo-SAX observations pointed on
the poor clusters \scsette and \scnove, placed in the A3558 complex at
the expected position of the shock front.
\\
The plan of the paper is the following.
In Sect. 2 we describe the characteristics of the A3558 complex and in 
Sect. 3 we give details about observations and data reduction.
In Sect. 4 and 5 we present the analysis of \scsette and \scnove, respectively;
in Sect. 6 we look for the possible existence of a soft excess in the 
Beppo-SAX LECS data, while the analysis of Beppo-SAX PDS data is given 
in Sect. 7. Finally, in Sect. 8 we discuss and summarize the results.

%------------------------------------------------------------------------------
% FIGURE 1 
\begin{figure}
\centering
\includegraphics[width=\hsize]{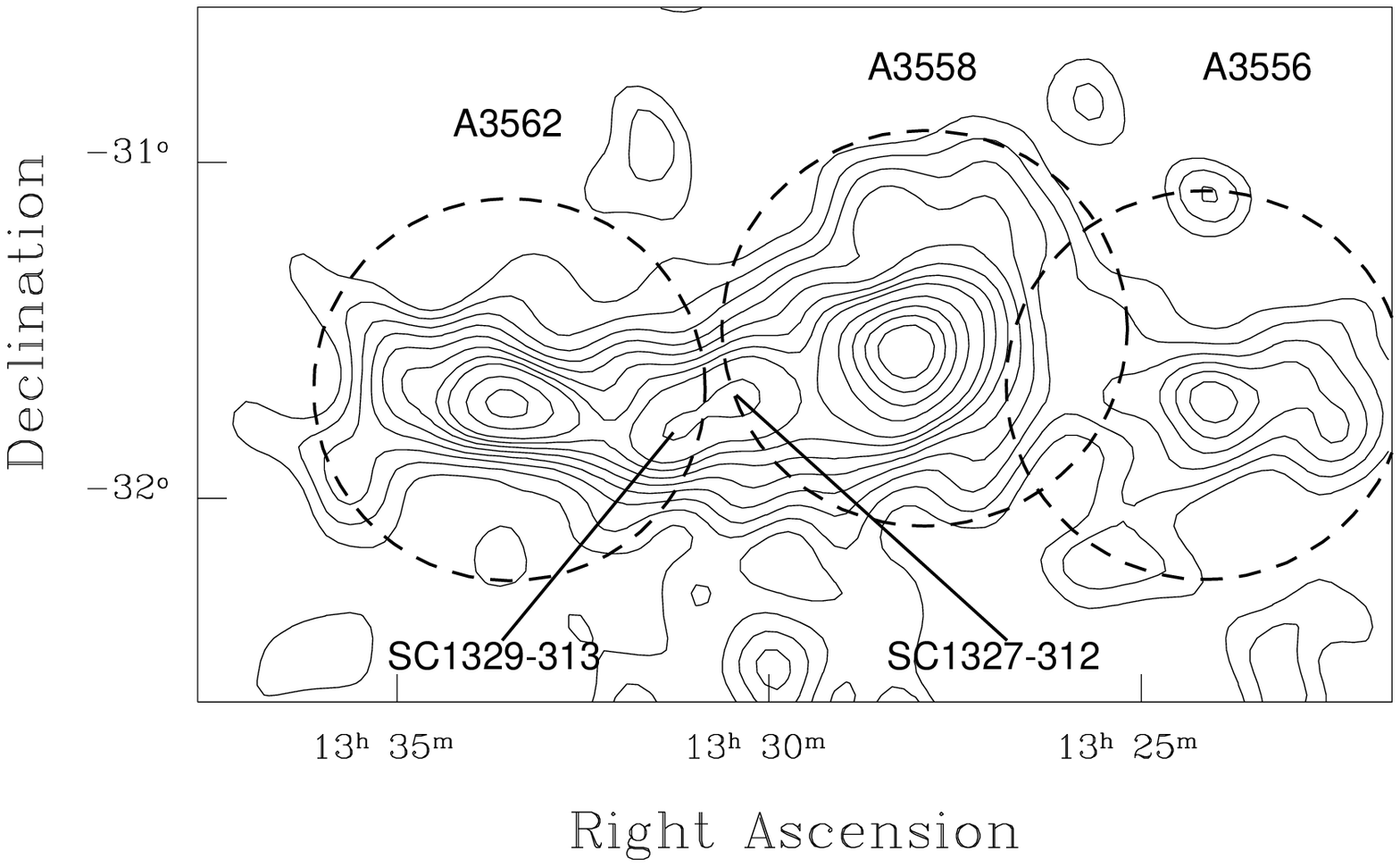} 
\includegraphics[width=\hsize]{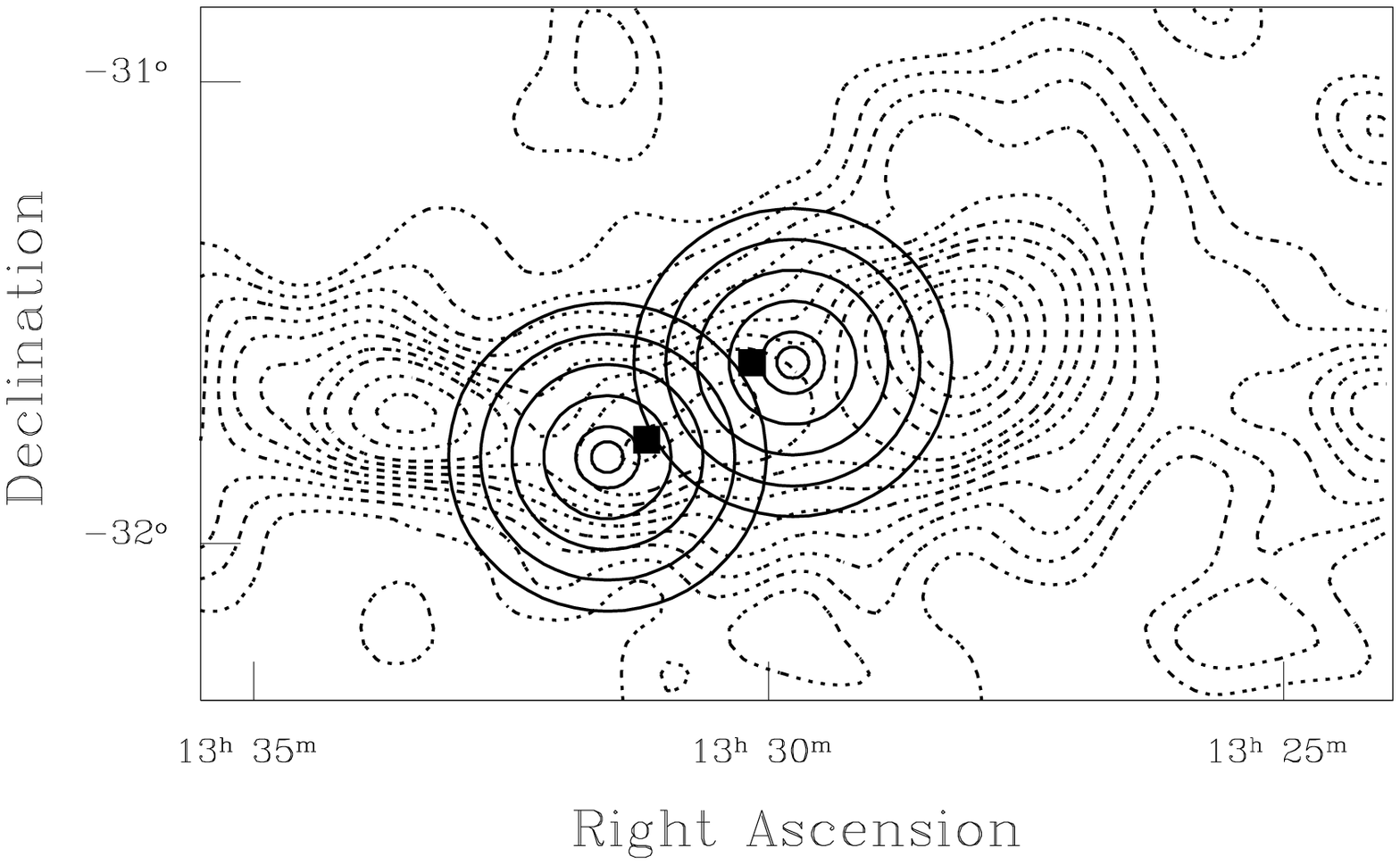} 
\caption{ Upper panel: optical galaxy isodensity contours of the A3558 cluster 
complex. The data have been smoothed with a Gaussian of 6 arcmin of FWHM.
Circles of 1 Abell radius have been drawn as dashed curves around each
cluster and the optical position of \scsette and \scnove is indicated.
Lower panel: close up of the same region, with superimposed the Beppo-SAX
fields. Concentric annuli correspond to the regions analyzed in Sect. 4.2
and 5.2. Solid squares indicate the position of the optical three-dimensional
substructures associated to \scsette (T478 on the right) and to \scnove (T520 
on the left), found by Bardelli et al. (\cite{bardelli98b}); 
see the text for more details.
} 
\label{fig:iso}
\end{figure}
%-------------------------------------------------------------------------------

%
%-------------------------------------------------------------------------------
%
\section{The A3558 cluster complex}

The A3558 complex (Bardelli et al. \cite{bardelli94}) is a chain of clusters 
almost perpendicular to the line of sight, spanning $\sim 7$ \hmpc at a 
redshift of $\sim 0.05$, and is 
formed by three ACO clusters (A3558, A3562 and A3556) and two poor groups 
(dubbed \scsette and \scnove by Breen et al. \cite{breen94}). 
This structure represents a density excess in number of galaxies of 
$N/\bar{N} \sim 45$, comprising a total mass in the range $\sim 1-5 \times 
10^{15}$ h$^{-1}$ M$_{\odot}$ (Bardelli et al. \cite{bardelli00}). 
A simple dynamical model shows that, with these characteristics, the complex 
is expected to be in the late stage of collapse. 
\\
From the two-dimensional distribution of optical galaxies (Figure \ref{fig:iso})
and a redshift survey of $\sim 700$ galaxies, we found that the external 
objects seem to form an envelope sourronding the cluster cores  
(Bardelli et al. \cite{bardelli94}, \cite{bardelli98a}). 
Detailed substructure analysis (Bardelli et al. \cite{bardelli98b}) revealed 
a large number of subclumps, suggesting that the complex is far from a relaxed 
state.
\\     
Considering the X-ray band, Bardelli et al. (\cite{bardelli96}) and 
Kull \& B\"ohringer (\cite{kull99}) showed that the whole chain is embedded 
in a hot gas filament.
The clusters of the chain have been extensively studied as single entities  
(Bardelli et al. \cite{bardelli96}, Ettori et al. \cite{ettori97}, 
Markevitch et al. \cite{markevitch98}, Hanami et al. \cite{hanami99},
Ettori et al. \cite{ettori00}), finding evidences of disturbance. 
\\
Remarkable results are found also in the radio band (Venturi et al. 
\cite{venturi00}): 
comparing the bivariate radio-optical luminosity function of radiogalaxies
in this complex with that of radiogalaxies in a sample of ``normal" clusters 
(Ledlow \& Owen \cite{ledlow96}), we found a significant lack of radiosources. 
It seems that the cluster interaction ``switched off" the central engines 
of the AGNs. Moreover, a relic of radiogalaxy has been found at the edge of 
a probably merging event between A3556 and a smaller group, projected along 
the line of sight (Venturi et al. \cite{venturi98}).
\\
In order to explain these facts, we adopted the working hypothesis that
the A3558 complex is the remanant of a cluster-cluster collision seen just
after the first core-core encounter
(Bardelli et al. \cite{bardelli98b}). 
We speculated that a cluster 
collided with A3558 and its remnants are visible as the overdensity regions of 
\scsette, \scnove and A3562. In this framework, A3556 would be formed by the 
members of the intervening cluster remained in the backside part of the 
merging direction.
Therefore, all galaxies outside A3558 would belong to the destroyed cluster
and form the clumpiness expected from the simulations.
\\
Hydrodynamical simulations have shown that mergers usually produce shocks in 
the intracluster medium (see e.g. Sarazin \cite{sarazin00}, Roettiger et al. 
\cite{roettiger97}, Takizawa \& Mineshige \cite{takizawa98}, Ricker \& Sarazin
\cite{ricker01}): at early stages, the shocked region
is located between the nuclei of the impacting clusters, while at later
times the shocks sweep over the centers and reach the outer regions.
\\ 
In order to completely understand which is the chain of events that
created the A3558 complex and in particular the time scale of the merging,
it is important to try to individuate the locus of the shock (if present). 
The most natural place to be explored is the region between A3562 and A3558,
where the two poor clusters \scsette and \scnove are located.

%
%-------------------------------------------------------------------------------
%
\section{Observations and data reduction}

The clusters \scsette and \scnove were observed by the Beppo-SAX
satellite (Boella et al. \cite{boella97a}) in the periods 1999 December 
28-30 and 2000 January 15-18, respectively.
We discuss here the data from two of the instruments onboard Beppo-SAX: the 
Medium-Energy Concentrator Spectrometer (MECS) and the Low-Energy Concentrator
Spectrometer (LECS). The MECS (Boella et al. \cite{boella97b}) is presently
composed of two units, working in the [1--10] keV energy range. At
6 keV, the energy resolution is $\sim 8\%$ and the angular resolution
is $\sim 0.7'$ (FWHM). The LECS (Parmar et al. \cite{parmar97}),
consists of an imaging X-ray detector, working in the [0.1--9] keV
energy range, with 20$\%$ spectral resolution and $0.8'$
(FWHM) angular resolution (both computed at 1 keV). Standard
reduction procedures and screening criteria have been adopted to
produce 
linearized (i.e. corrected for intrinsic distorsion of the detector),
cleaned (filtered to remove non-scientific event and to correct gain variations)
and equalized (in order to report the two MECS at the same energy scale)
event files.
The MECS (LECS) data preparation and linearization was performed using the 
{\sc Saxdas} ({\sc Saxledas}) package under {\sc Ftools} environment.
\\
We have taken into account the PSF-induced spectral distortions 
(D'Acri et al. \cite{dacri98}) in the MECS analysis using effective area files
produced with the {\it effarea} program.
All MECS spectra have been background subtracted using spectra
extracted from blank sky event files in the same region of the
detector as the source (see Fiore et al. \cite{fiore99}).
A detailed explanation of the MECS analysis is given in De Grandi \& Molendi 
(\cite{degrandi01}): in the following we'll concentrate on the most important 
steps. 
As done in Ettori et al. (\cite{ettori00}), for the LECS we have used two
redistribution matrices and ancillary response files, the first computed
for an on-axis pointlike source and the second for a source with a
flat brightness profile.  The temperatures and abundances we derive in
the two cases do not differ significantly, as the telescope vignetting
in the [0.1--4.0] keV energy range is not strongly dependent upon energy.
All spectral fits have been performed using XSPEC Ver. 10.00.
\\
The observation log is reported in Table \ref{tab:obs} and the field position
is shown in the lower panel of Figure \ref{fig:iso}. 
The observed count--rates for \scsette and for \scnove for the 2 MECS units
and within the central 8 arcmin (corresponding to 0.33 \hmpc) 
are 0.122 cts s$^{-1}$ and 0.060 cts s$^{-1}$, respectively. 
For the LECS data, the count--rates in the same region are 0.087 and 0.045,
respectively.

%------------------------------------------------------------------------------
% FIGURE 2 
\begin{figure}
\centering
\includegraphics[width=\hsize]{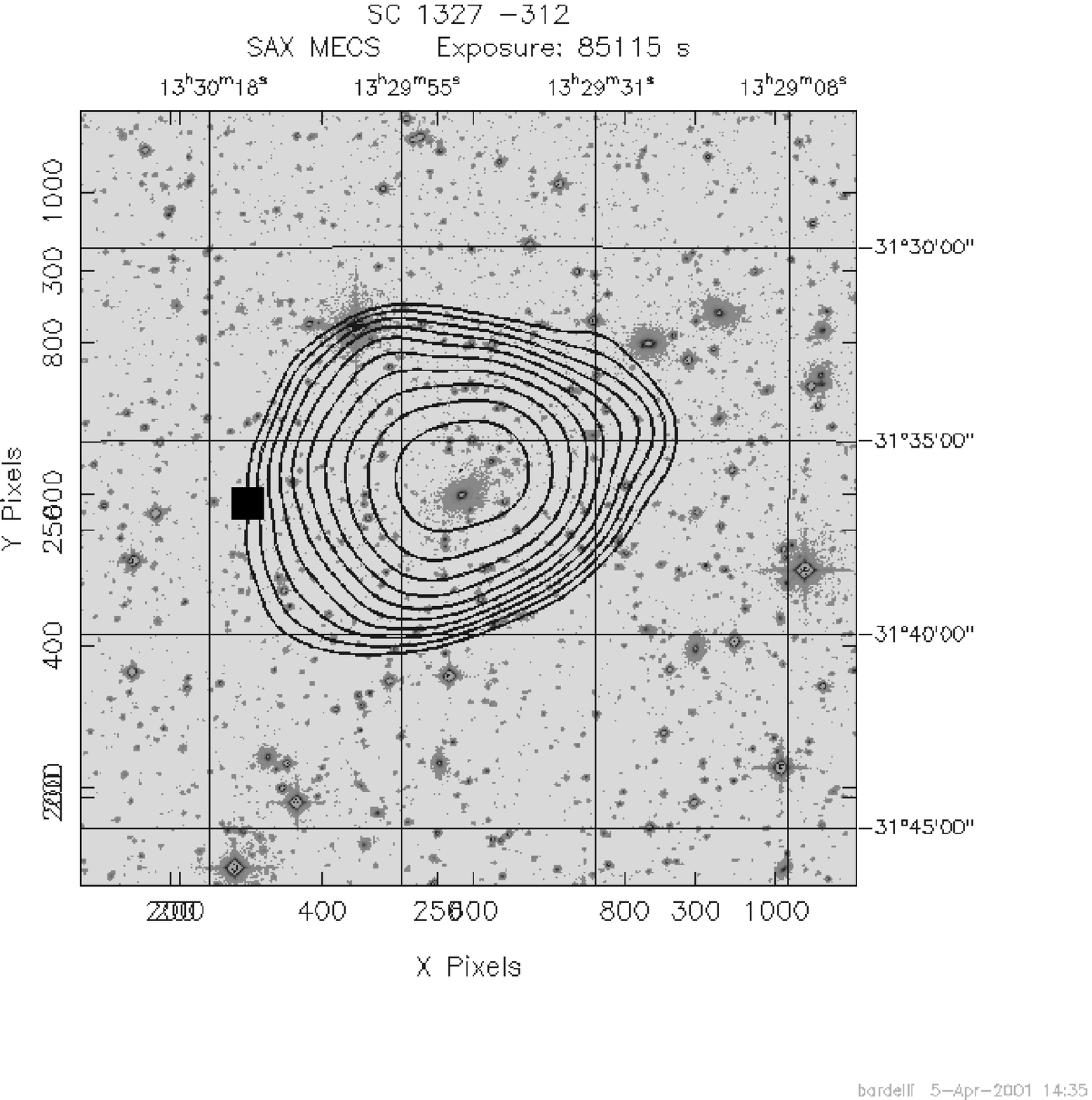} 
\caption{MECS image of \scsette in the [2--10] keV energy range. The data 
have been smoothed with a Gaussian of 6 pixels, corresponding to 0.8 arcmin.
In this figure we show the X-ray contours from the central region of 8 arcmin 
radius, superimposed to the Digital Sky Survey. 
The solid square indicates the position of the optical group T478 
(see the caption of Figure \ref{fig:iso} and the text for more details). }
\label{fig:smooima1327}
\end{figure}
%------------------------------------------------------------------------------

%-------------------------------------------------------------------------------
% TABLE 1.
\begin{table*}
\caption[]{ Beppo-SAX observation log (data are referred to MECS)}
\begin{flushleft}
\begin{tabular}{rrrrrr}
\hline\noalign{\smallskip}
Target & $\alpha$ (2000) & $\delta$ (2000) & date & exp.time & count--rate \\
  & $^h$ $^m$ $^s$  & $^o$ $'$ $''$  &   &  ksec  &    cts/s   \\
\noalign{\smallskip}
\hline\noalign{\smallskip}
\scsette  & 13 29 47 & -31 36 29 & 1999 Dec 28-30 & ~85.1 & 0.122 \\
\scnove   & 13 31 36 & -31 48 46 & 2000 Jan 15-18 & 130.2 & 0.060 \\
\noalign{\smallskip}
\hline
\end{tabular}
\end{flushleft}
\label{tab:obs}
\end{table*}
%-------------------------------------------------------------------------------

%
%-------------------------------------------------------------------------------
%
\section{\scsette}

The cluster \scsette ($\alpha_{2000}=13^h 29^m 47^s$, $\delta_{2000}=
-31^o 36' 29''$, from Bardelli et al. \cite{bardelli96}) is located 
23.6 arcmin ($\sim 1$ \hmpc) from the center of the dominant cluster A3558. 
In Figure \ref{fig:smooima1327}
the isodensity contours of the MECS image (in the energy range [2--10] keV)
are reported, after having applied a smoothing of 6 pixels, corresponding 
to 0.8 arcmin. 
The X-ray emission from the central 8 arcmin is overplotted 
on the optical data taken from the Digital Sky Survey:
the emission seems to be symmetric and centered 
on the brightest galaxy, which has $b_J=15.6$ and $v=15121$ km s$^{-1}$.
The distribution of optical galaxies does not show a clear overdensity at the 
position of this group: the nearest group found by the three-dimensional
substructure analysis of Bardelli et al. (\cite{bardelli98b}), dubbed T478, 
is $\sim 5$ arcmin away from the X-ray position, in the East direction
(see lower panel of Figure \ref{fig:iso} and Figure \ref{fig:smooima1327}).  
This offset could be due to the contamination of galaxies from the nearby 
cluster A3558, which decreases the significance of the optical overdensity.
Moreover the optical isodensity distribution in this region shows a dependence 
on the magnitude range (see figure 14 of Bardelli et al. \cite{bardelli94}), 
another indication of possible contamination from nearby structures. 
\\
This group appears in a ROSAT-PSPC observation pointed on A3558:
unfortunately, is has been observed off-axis, where the spatial resolution 
is strongly degraded; moreover its emission is disturbed by the
presence of a supporting rib of the detector (see Bardelli et al. 
\cite{bardelli96}). 
For this reason, we fitted an elliptical King model (i.e. with two core radii,
see Bardelli et al. \cite{bardelli96}) directly to 
the Beppo-SAX MECS image.
\\
We found that the best fit within a region of 8 arcmin radius is 
$r_{c1}=0.101$ \hmpc, $r_{c2}=0.135$ \hmpc and $\beta=0.492$.
Note that the core radius values correspond to $\sim 4$ times the
resolution and could be considered an unbiassed measure.  
 \\
In order to explore the dependence of the results on the PSF,
we deconvolved the image with a Clark Clean Method, finding that
the parameters vary by a 5$\%$, in the sense of reducing the core radii.  
\\ 
The only value present in the literature is $r_{c}=0.17\pm 0.01$ \hmpc
obtained by Breen et al. (\cite{breen94}) on Einstein-IPC data, fixing the 
slope to $\beta=0.6$.
With our parameters 
and using the estimated global temperature of the
hot gas (see below),
the total mass is 
$M(<1$\hmpc$) \sim 2.0\times 10^{14}$ h$^{-1}$ M$_{\odot}$, with a baryonic 
fraction of $\sim 7\%$ h$^{-1.5}$.
\\ 
The luminosity
resulted to be $3.76 \times 10^{43}$ h$^{-2}$ erg s$^{-1}$ in the [2--10] keV
band and $1.77 \times 10^{43}$ h$^{-2}$ erg s$^{-1}$  in the [0.6--3] keV
band. These values have been estimated within a distance of $0.5$ \hmpc from the
cluster center.  

%-------------------------------------------------------------------------------
% FIGURE 3 
\begin{figure}
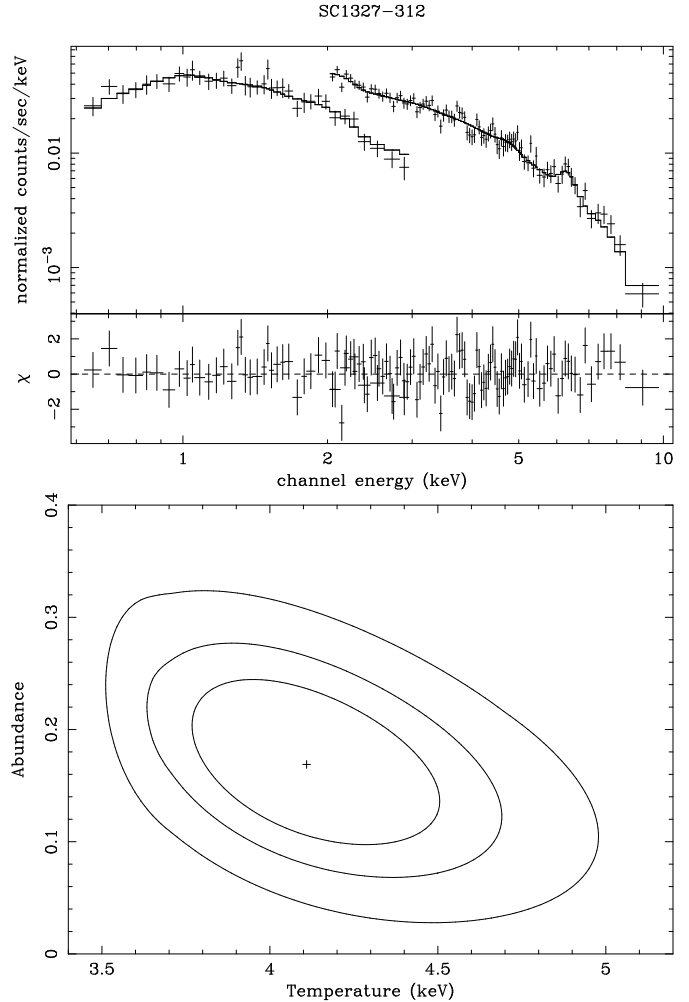

\centering
\includegraphics[angle=-90,width=\hsize]{paper_sc_fig3a.ps}
\includegraphics[angle=-90,width=\hsize]{paper_sc_fig3b.ps}
\caption{Spectrum and confidence ellipse of the estimated temperature
and abundance on MECS$+$LECS data within 8 arcmin from the center of \scsette.}
\label{fig:sc1327spec}
\end{figure}
%-------------------------------------------------------------------------------

%
%-------------------------------------------------------------------------------
%
\subsection{Global temperature and abundance}

In order to find the global temperature, we extracted a circular region
of 8 arcmin from the center of the MECS data in the range [2--10] keV. 
We fitted the spectrum with a {\it mekal} (Mewe et al. \cite{mewe95}, 
Kaastra \cite{kaastra92}) model with an absorbing Galactic hydrogen column 
({\it wabs} model), as implemented in the XSPEC package.     
After having checked that the fitted Galactic absorption is consistent
with the literature measurement of $3.88\times 10^{20}$ cm$^{-2}$ 
(Dickey \& Lockman \cite{dickey90}), we fixed it to this latter value.  
The results are reported in Table \ref{tab:global}. 
\\
In order to use all the information present in the Beppo-SAX observation, 
we added the data of the LECS instrument in the [0.6--3] keV band.  
From a ROSAT-PSPC image, Bardelli et al. (\cite{bardelli96}) found that 
superimposed to the cluster image (at $\sim 4.6$ arcmin from the cluster 
center) there is a K1 III star (SAO 204527 or HD 117310), whose
X-ray emission is well fitted by a Raymond-Smith model with $kT=1.10$ keV.  
This emission is likely affecting only the LECS energy range: for this reason
we fitted the {\it mekal} model adding a Raymond-Smith model with 
$kT=1.10$ keV and abundance of $0.20$. The resulting temperature is 
$4.11^{+0.43}_{-0.36}$ keV and the abundance is $0.17^{+0.08}_{-0.08}$,
with a reduced $\chi^2$ of 1.03 with 189 degrees of freedom.
However, neglecting the star contribution, the results are very similar,  
with a reduced $\chi^2$ of 1.05 with 192 degrees of freedom.    
Also these results are reported in Table \ref{tab:global}.
\\
In Figure \ref{fig:sc1327spec} we show the combined LECS$+$MECS spectrum of 
\scsette, overplotted to the fit, and the corresponding confidence ellipse 
of the temperature and abundance parameters. 
The value of the temperature is in agreement with the determination of 
$kT=3.85$ keV from ROSAT-PSPC (Bardelli et al. \cite{bardelli96}) and 
is consistent at 1.5 sigma with the ASCA determinations of $kT=3.76 \pm 0.13$ 
keV (Hanami et al. \cite{hanami99}). 
\\
Following the $\sigma-T$ relation of Lubin \& Bahcall (\cite{lubin93})
[$\sigma=332(kT)^{0.6}$ km s$^{-1}$], the determined temperature 
implies a velocity dispersion of $775^{+47}_{-41}$ km s$^{-1}$,
consistent with the value of Bardelli et al. (\cite{bardelli98a}) 
of $691^{+158}_{-246}$ km s$^{-1}$.

%-------------------------------------------------------------------------------
% TABLE 2.
\begin{table*}
\caption[]{ Spectral results from a region of 8 arcmin from the center
of \scsette and \scnove. Errors are $90\%$ confidence level.} 
%\begin{flushleft}
\begin{tabular}{rrrr}
\hline\noalign{\smallskip}
Object & $kT$ (keV) & Abundance & Reduced $\chi^2$  (d.o.f) \\
\noalign{\smallskip}
\hline\noalign{\smallskip}
\scsette MECS  & $3.93^{+0.16}_{-0.17}$ & $0.17^{+0.05}_{-0.04}$ & 1.05 
 (133) \\
\noalign{\smallskip}
\scsette LECS+MECS  & $4.11^{+0.43}_{-0.36}$ & $0.17^{+0.08}_{-0.08}$ & 1.03
 (189) \\ 
\noalign{\smallskip}
\scnove MECS  & $3.43^{+0.28}_{-0.25}$ & $0.22^{+0.11}_{-0.11}$ & 1.02 
 (125) \\ 
\noalign{\smallskip}
\scnove LECS+MECS  & $3.49^{+0.27}_{-0.24}$ & $0.21^{+0.09}_{-0.11}$ & 0.92
 (181) \\ 
\noalign{\smallskip}
\hline
\end{tabular}
%\end{flushleft}
\label{tab:global}
\end{table*}
%-------------------------------------------------------------------------------

%
%-------------------------------------------------------------------------------
%
\subsection{MECS spatially resolved spectroscopy}

\begin{figure}
\centering
\includegraphics[width=\hsize]{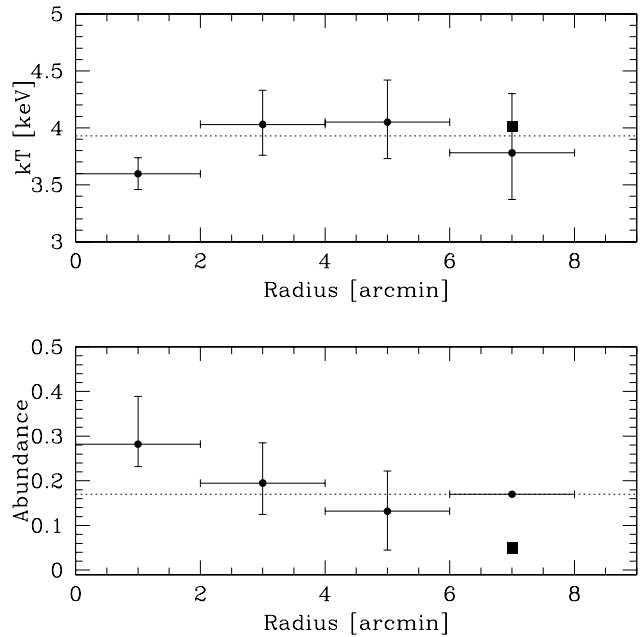} 
\caption{Temperature and abundance radial profile of \scsette. 
The vertical bars correspond to $68\%$ errors, while the horizontal
bars represent the bins used to extract the counts. Dotted lines indicate 
the global fit values. Squares are values obtained by fixing the 
abundance value at the linear extrapolation of the abundance profile.}
\label{fig:radprofSC1327}
\end{figure}

\begin{figure}
\centering
\includegraphics[width=\hsize]{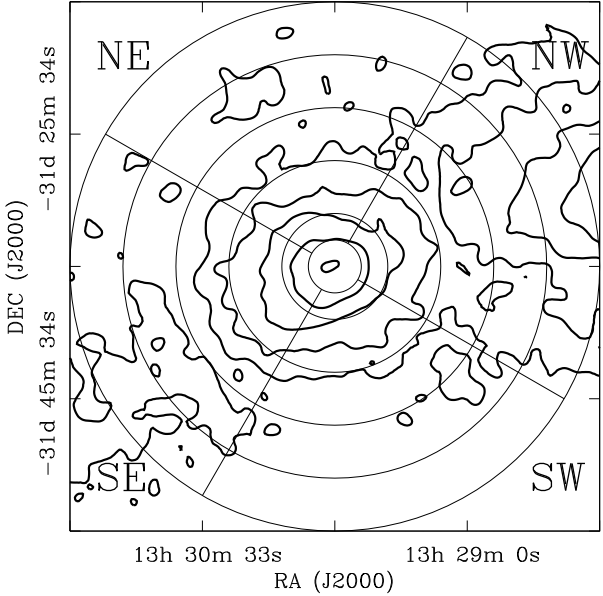} 
\caption{ Beppo-SAX MECS image of \scsette. Contour levels (thick lines) are 
spaced logarithmically. Overlapped to the contours the concentric circles and
the cross, drawn as thinner lines, show how the cluster has been
divided to obtain temperature maps. The position angle (computed from
North to East) is $60^o$. The NW sector is pointing toward A3558 and the SE 
one toward \scnove. Note that the effect of the supporting structure of
the MECS entrance window (i.e. the strongback) has been not corrected
in this images. }
\label{fig:sc1327secmap}
\end{figure}

The cluster emission in the central 8 arcmin circle has been divided into 
$2'$ wide concentric annuli, centered on the X-ray emission peak. 
To all spectra accumulated from these annular regions, we have applied
a single temperature model (i.e. {\it mekal} model) absorbed for the
nominal Galactic hydrogen column density ({\it wabs} model) to derive
temperature and metal abundances. The adopted column density for
\scsette is $3.88\times 10^{20}$ cm$^{-2}$ (see Sect. 4.1), and the 
energy range considered for the spectral fitting procedure is [2--10] keV.
\\
In Figure \ref{fig:radprofSC1327} we report the temperature and abundance 
profiles of \scsette in annuli around the cluster center. The vertical
bars correspond to the $68\%$ errors and the horizontal bars represent the 
bins used to extract the counts. For the
last point, for which the abundance is not constrainted, we fixed it at the
value of the global fit, i.e. 0.17. For this reason this last abundance point
is plotted without error bars.
If we extrapolate the abundance profile with a linear fit (see below), we
obtain the value indicated with the square: in the temperature plot, the 
square indicates the $kT$ derived using this abundance value.
These points were plotted in order to give a feeling of the dependence 
of the fitted temperature on the abundance determination.    
Finally, dotted lines correspond to the values obtained from the global fit.
\\ 
The temperature profile appears to be constant, a part a weak indication 
of a decrease in the most central bin: averaging the values of all bins 
we find $kT=3.88_{-0.16}^{+0.15}$ keV, well in agreement with the global
fitted temperature. 
On the contrary, the abundance seems to present a decreasing trend. In order 
to extrapolate the trend to the $6-8$ arcmin bin, we fitted the linear 
relation $(0.33-0.041 r)$. 
\\
We have performed two-dimensional spectral analysis of \scsette by
extracting spectra from sectors of annuli as shown in
Figure \ref{fig:sc1327secmap}: out to 4 arcmin each annulus is $2'$ wide, 
beyond this radius the annuli are $4'$ wide.
This figure shows the MECS image of the cluster with the four sectors
overlaid: 
note that the contours shown in Figure \ref{fig:smooima1327} 
correspond only to the inner part of this plot. 
We have tilted the sectors with a position angle (measured
from North to East) of 60 degrees in order to have the North-West
sector pointing toward A3558 and the South-East one toward \scnove.  
The considered energy band is again [2--10] keV, except for
the $8' - 12'$ annulus. In this annulus we have applied a
correction for the absorption caused by the strongback supporting of the
detector window and in this case the energy range is [3.5--10] keV to avoid 
the low energy part of the spectrum, where the correction for the 
strongback is less reliable.
%
%-------------------------------------------------------------------------------
% FIGURE 4 
%-------------------------------------------------------------------------------
%
%------------------------------------------------------------------------------
% FIGURE 5 
%-------------------------------------------------------------------------------
%
%-------------------------------------------------------------------------------
% FIGURE 6 
\begin{figure}
\centering
\includegraphics[width=\hsize]{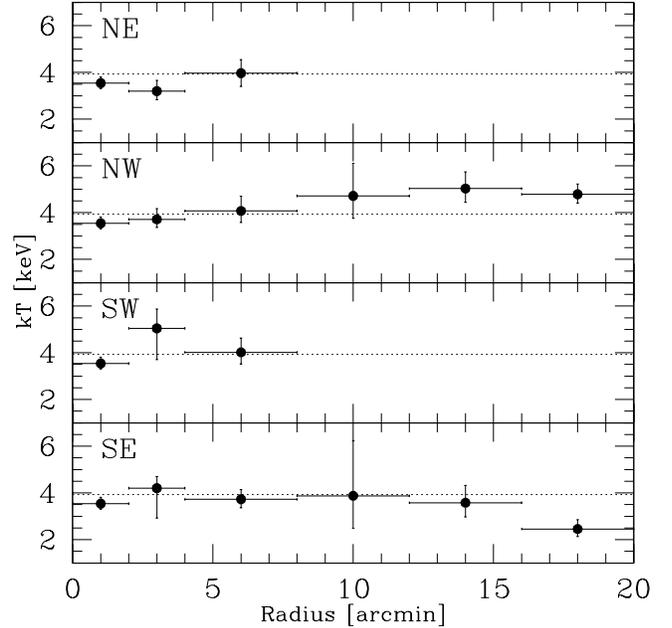}
\caption{Two-dimensional radial profile of the temperature of \scsette.
The four panels show the results of the four sectors indicated in
Figure \ref{fig:sc1327secmap}. 
The vertical bars correspond to one sigma errors, while the horizontal
bars represent the bins used to extract the counts. Dotted lines indicate 
the global fit values.
}
\label{fig:radprofmapSC1327}
\end{figure}
%-------------------------------------------------------------------------------
\\
In Figure \ref{fig:radprofmapSC1327} we show the temperature profiles derived
for \scsette from the spectral fits for each of the 4 sectors. In all profiles
we have included the temperature obtained for the central circular
region with radius $2'$.
The radial temperature profile for each sector stops at the last
annulus where the source counts are more than $30\%$ of the total
(i.e. source plus background) counts.
This is the reason why there are different radial extensions for the
profiles in the four quadrants: in the outermost bins in the North-West
and South-East sectors there is a contribution from the emission of the 
adjacent clusters.
\\
More specifically, sector North-West points toward A3558, a cluster at
a global temperature of 5.5 keV (Markevitch \& Vikhlinin \cite{markevitch97}), 
which is about 24 arcmin away from \scsette (note that our measurements
extend up to 20 arcmin from the cluster center only). Markevitch \&
Vikhlinin (\cite{markevitch97}) measured the temperature map for A3558 
using ASCA data: they find a temperature of about $4.0\pm0.8$ keV ($90\%$
c.l. errors) in their sector of annulus pointing towards \scsette
(i.e. region 9 in figure 2 of Markevitch \& Vikhlinin \cite{markevitch97}), 
in agreement with the mean temperature we find for the three outermost
bins of sector North-West, i.e. $4.7\pm0.6$ keV ($90\%$ c.l.).
We have fitted the temperature in sector North-West with a constant $kT$ model,
finding a mean temperature of $4.0\pm0.2$ with a reduced $\chi^2=1.6$;
we find a marginal (at $\sim 98\%$ c.l. using the F-test) evidence of
temperature increase from 0 to 20 arcmin by fitting the data with a
linear model. At the scales considered here we can exclude strong
temperature enhancements along this direction.

%
%-------------------------------------------------------------------------------
%
\section{\scnove}

\begin{figure}
\centering
\includegraphics[width=\hsize]{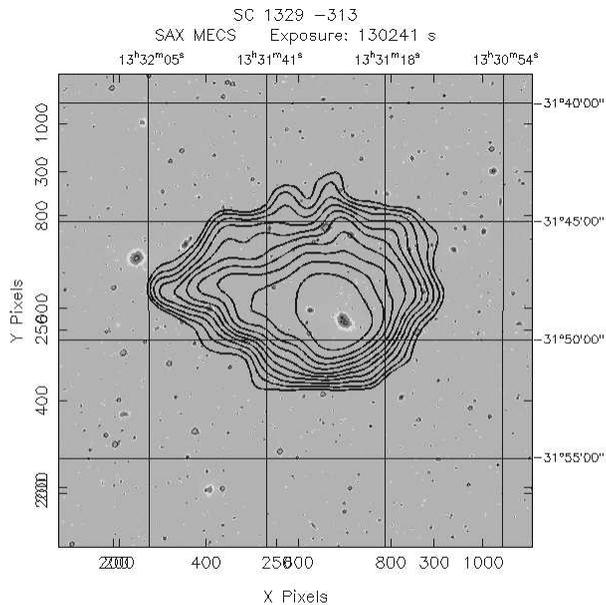} 
\caption{MECS image of SC1329-313 in the [2--10] keV energy range. The data 
have been smoothed with a Gaussian of 6 pixels, corresponding to 0.8 armin.
In this figure we show the X-ray contours from the central region of 8 arcmin 
radius, superimposed to the Digital Sky Survey. }
\label{fig:smooima1329}
\end{figure}

The cluster \scnove ($\alpha_{2000}=13^h 31^m 36^s$, $\delta_{2000}=
-31^o 48' 46''$, from Breen et al. \cite{breen94}) is located 27.1 arcmin 
($\sim 1.2$ \hmpc) from the center of the  cluster A3562 and 26.3 arcmin 
from \scsette.
In Figure \ref{fig:smooima1329}
the isodensity contours of the MECS image (in the energy range [2--10] keV)
are reported, after having applied a smoothing of 6 pixels, corresponding 
to 0.8 arcmin. 
The emission from the central 8 arcmin is overplotted
to the Digital Sky Survey optical image.
\\
The isodensity contours of \scnove appear elongated with a tail pointing toward 
A3562. The brightest galaxy ($b_J=15.6$ and $v=12928$ km s$^{-1}$) appears
hosting a double nucleus. 
Fitting an elliptical King model, we find $r_{c1}=0.146$ \hmpc, 
$r_{c2}=0.185$ \hmpc and $\beta=0.499$.
We find also an excess of emission at $\sim 1.7$ arcmin from the
fitted center, roughly in correspondence of the brightest galaxy. 
\\
Also in this case, the Clark Clean Method deconvolution of the image leads to
neglegible variations of fitted parameters.   
Using these values and the global temperature (see below), we estimate
a mass of
$M(<1$ \hmpc$) \sim 5.7 \times 10^{14}$ h$^{-1}$ M$_{\odot}$,  
with a gas mass fraction of $\sim 8\%$ h$^{-1.5}$.
\\
The [2--10] keV luminosity is $2.36 \times 10^{43}$ h$^{-2}$ erg s$^{-1}$
and the emission in the [0.6-3] keV band is $1.17 \times 10^{43}$ h$^{-2}$ 
erg s$^{-1}$  within $0.5$ \hmpc from the center.

%------------------------------------------------------------------------------
% FIGURE 7  
%-------------------------------------------------------------------------------

%-------------------------------------------------------------------------------
% FIGURE 8 
\begin{figure}
\centering
\includegraphics[angle=-90,width=\hsize]{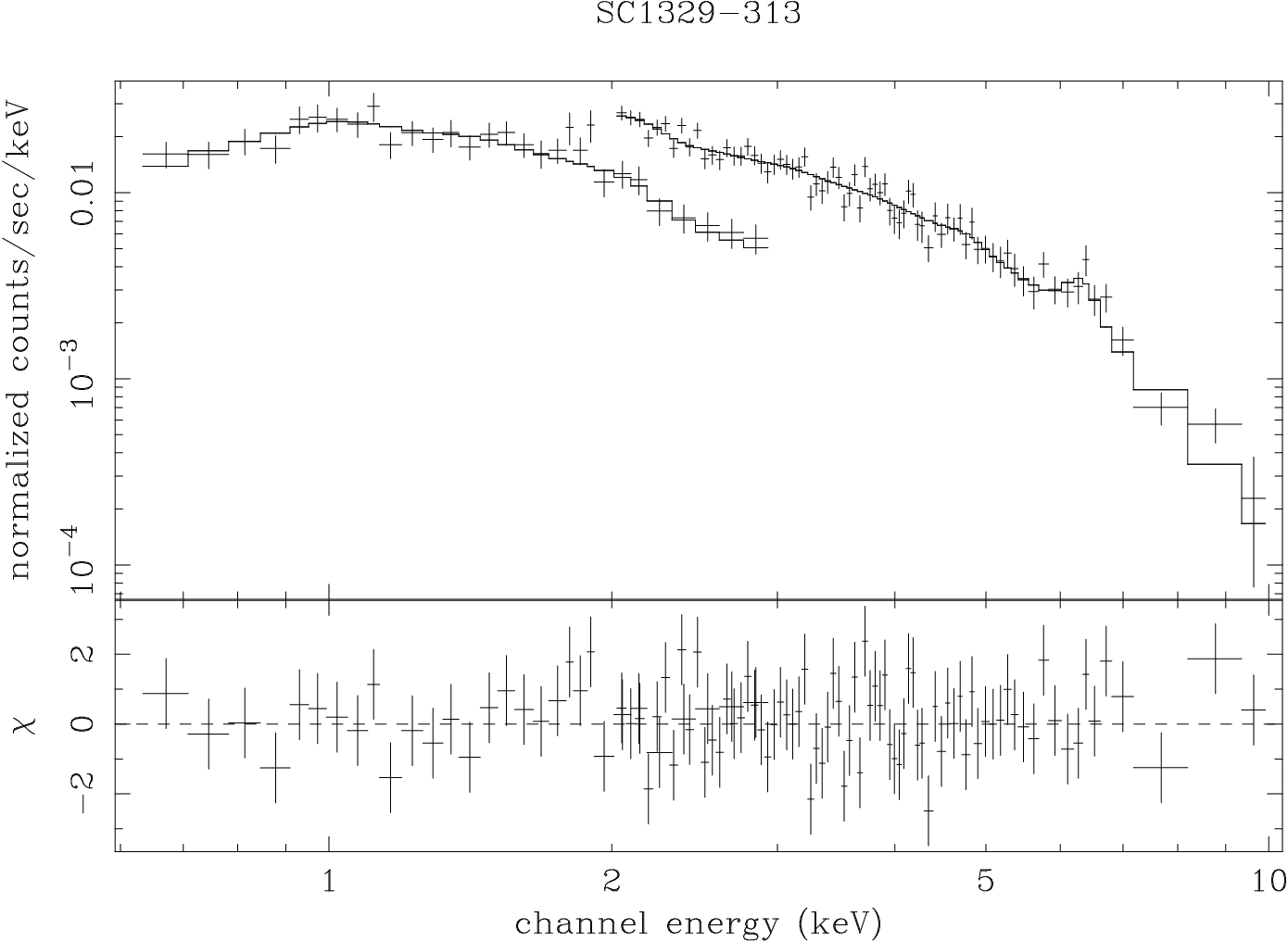}
\includegraphics[angle=-90,width=\hsize]{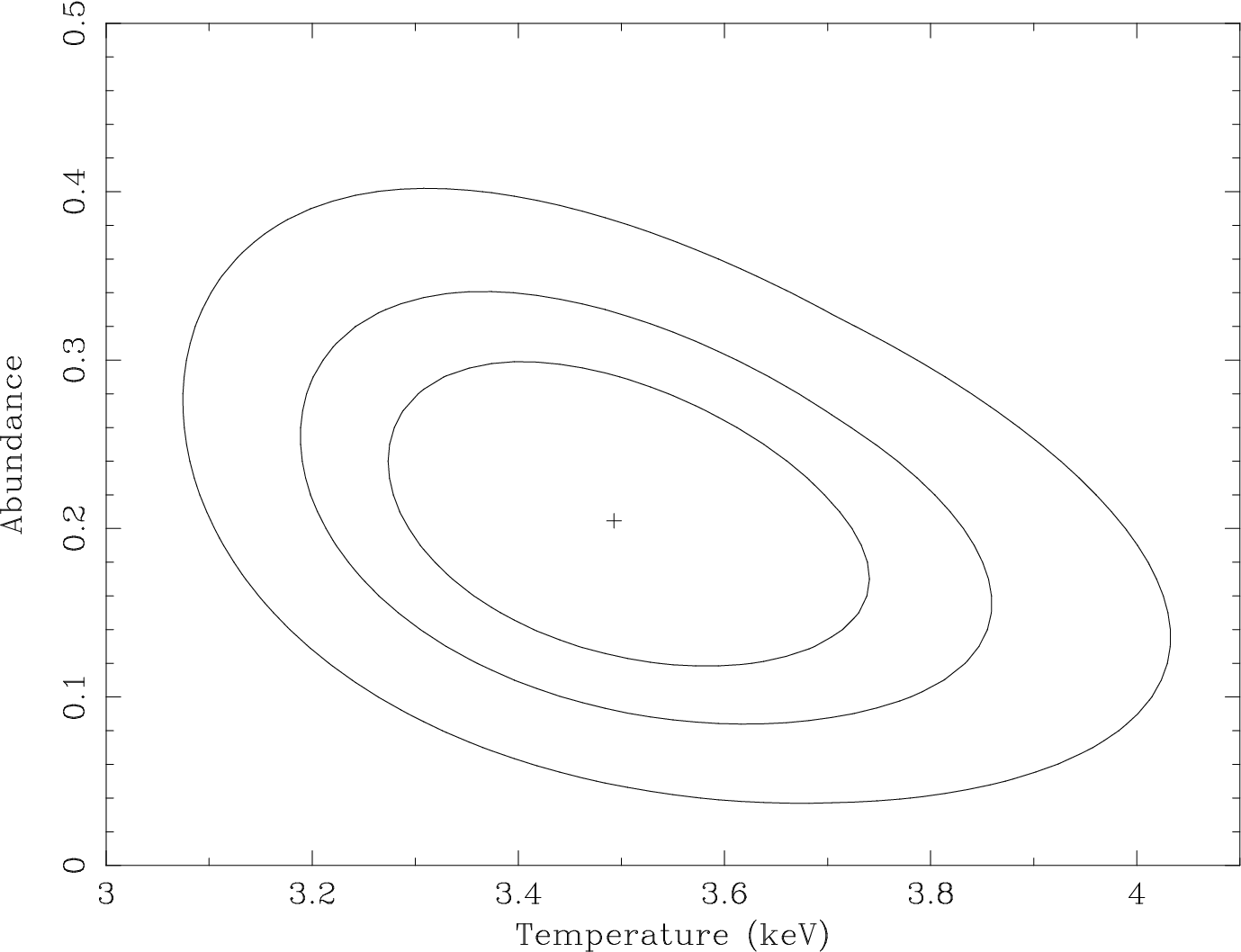}
\caption{Spectrum and confidence ellipse of the estimated temperature
and abundance on MECS$+$LECS data within 8 arcmin from the center of \scnove.}
\label{fig:sc1329spec}
\end{figure}
%------------------------------------------------------------------------------

%-------------------------------------------------------------------------------
% FIGURE 9 
\begin{figure}
\centering
\includegraphics[width=\hsize]{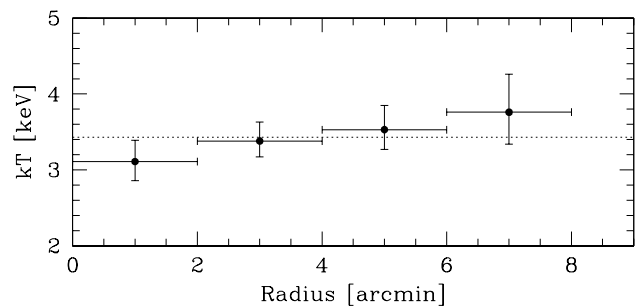} 
\caption{Temperature radial profile of \scnove. The vertical bars correspond 
to $68\%$ 
errors, while the horizontal bars represent the bins used to extract the 
counts. Dotted line corresponds to the global fit.}
\label{fig:radprofSC1329}
\end{figure}
%-------------------------------------------------------------------------------

%
%-------------------------------------------------------------------------------
%
\subsection{Global temperature and abundance}

The global results from the region of 8 arcmin radius are 
$kT=3.49^{+0.27}_{-0.24}$ keV and an abundance of $0.21^{+0.09}_{-0.11}$, 
with a reduced $\chi^2$ of 0.92 with 181 degrees of freedom.
We fixed the hydrogen column to $3.98\times 10^{20}$ cm$^{-2}$ (Dickey \& 
Lockman \cite{dickey90}), after having checked that it is consistent with
the fitted value. 
Note that the addition of the LECS dataset does not significantly modify the 
temperature determination.
The results are reported in Table \ref{tab:global}. 
The derived temperature is significantly lower (at 3.3 sigma) than the ASCA
determination of $kT = 4.21^{+0.15}_{-0.23}$ keV (Hanami et al. 
\cite{hanami99}): it is unclear which can be the origin of this discrepancy,
given the fact that generally Beppo-SAX and ASCA results for global 
temperatures of clusters are in agreement (De Grandi \& Molendi 
\cite{degrandi02}).
\\
In Figure \ref{fig:sc1329spec} we show the LECS+MECS spectrum of \scnove, with 
overplotted the best fit, and the corresponding confidence ellipse 
for the temperature and abundance parameters. 
\\
The implied velocity dispersion is $702^{+33}_{-49}$ km s$^{-1}$.
From the substructure analysis of Bardelli et al. (\cite{bardelli98b}) 
it appeared
that this cluster has a bimodal velocity distribution, with the two peaks
separated by $\sim 1500$ km s$^{-1}$  ($\langle v \rangle =13280$ and 
$14960$ km s$^{-1}$) and superimposed along the line of sight.
It is difficult to determine which subclump is associated with this diffuse 
emission. The optical substructures have centers offset by $\sim 9 $ arcmin 
(the farthest, T496) and $\sim 6$ arcmin (the nearest, T520) westward. 
The velocity dispersions
are $482^{+87}_{-49}$ km s$^{-1}$ and $537^{+87}_{-32}$ km s$^{-1}$,
respectively, lower than the predicted one of 2.2 and 1.6 $\sigma$. 
However, considering that the brightest galaxies nearby the X-ray 
emission are at $\sim 13000$ km s$^{-1}$, the group responsible for the 
emission is presumably T520; its position is shown in the lower panel
of Figure \ref{fig:iso}.

%
%-------------------------------------------------------------------------------
%
\subsection{MECS spatially resolved spectroscopy}

\begin{figure}
\centering
\includegraphics[width=\hsize]{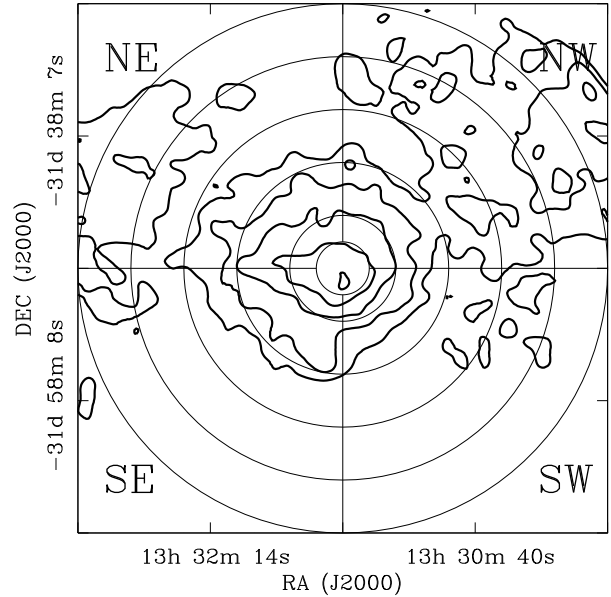} 
\caption{ Same as Figure \ref{fig:sc1327secmap} for \scnove. The position 
angle of the grid is zero. In this case the NW sector is pointing toward
\scsette and the NE one toward A3562. }
\label{fig:sc1329secmap}
\end{figure}

\begin{figure}
\centering
\includegraphics[width=\hsize]{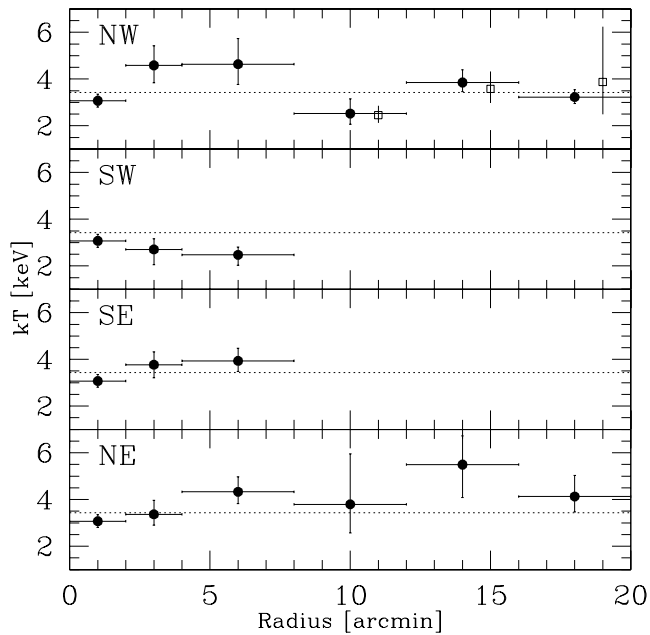}
\caption{Two-dimensional radial profile of the temperature of \scnove 
The four panels show the results of the four sectors indicated in
Figure \ref{fig:sc1329secmap}. 
The vertical bars correspond to one sigma errors, while the horizontal
bars represent the bins used to extract the counts. Dotted lines indicate 
the global fit values.
Open squares in the last bins of the NW sector indicate the temperature
estimated from the corresponding bins in the SE sector of \scsette 
(see text for details).  }
\label{fig:radprofmapSC1329}
\end{figure}

For this cluster we have used the same spatial binning for the concentric 
annuli and the same energy bands as for \scsette (see Sect. 4.2).  
Given the low statistics of counts, the abundance is not constrained:
for this reason we choose to fix it to the global value (i.e. 0.22) for all 
bins.  
In Figure \ref{fig:radprofSC1329}, the radial temperature profile of \scnove 
is shown. A very marginal indication of temperature increase seems to be 
present.
\\
In Figure \ref{fig:sc1329secmap} we plot the MECS image of \scnove with 
overlaid thefour sectors used for the two-dimensional spectral analysis:
again the contours shown in Figure \ref{fig:smooima1329} 
correspond only to the inner part of this plot. 
The position angle in this case is 0 degrees, with the North-West sector
pointing towards \scsette and the North-East one towards A3562.
Again, in the sectors where there is a contribution from the emission of 
nearby objects, it is possible to extend the analysis up to 20 arcmin.
\\
In Figure \ref{fig:radprofmapSC1329} we show the temperature profiles derived
for \scnove from the spectral fits for each of the 4 sectors. In all profiles
we have included the temperature obtained for the central circular
region with radius $2'$.
\\
We have fitted the temperature profile of sector North-West with a
constant model, finding a mean temperature of $3.4\pm0.2$ keV; the
temperature enhancements in the second (2-4 arcmin) e third (4-8
arcmin) bins are statistically significant at less than $1.5\sigma$.
\\
In sector North-East we find X-ray emission from the cluster A3562 which
is located at about 27 arcmin from the center of \scsette. Ettori
et al. (\cite{ettori00}) studied the temperature map of A3562, finding a
temperature of $3.9^{+1.1}_{-0.8}$ keV in the annulus 8-16 arcmin in the
sector toward \scsette (sector West in figure 9 of Ettori et al.).  
This temperature is in agreement with our result in annulus 12-20 arcmin 
(i.e. $3.5\pm0.3$ keV), which is the region where the two cluster emissions 
are overlapping each others.
\\ 
It appears that the increase of temperature is mainly due to the 
second and third bin of the NW sector (with temperatures of $\sim 4.5$ keV).
This could be an indication of the presence of heating of gas, which
added to the elongation of the isophotes to the other side of \scnove,
are signs of interaction between this group and \scsette. 
\\
Finally, given the proximity of the Beppo-SAX observations of the two poor 
clusters, the last three bins of the SE sector of \scsette correspond with 
great overlap to the last three bins of the NW sector of \scnove
(see lower panel of Figure \ref{fig:iso}). 
Given the fact that the two observations are statistically independent, 
this is an opportunity to test how robust are the temperature determinations 
at large radii. In Figure \ref{fig:radprofmapSC1329}, we plotted as 
squares the temperature determinations of \scsette in the corresponding bins,
with a small offset on the X axis for clarity. 
In all bins, the two temperature estimates are very well consistent with
each other.

%------------------------------------------------------------------------------
% FIGURE 10 
%------------------------------------------------------------------------------

%-------------------------------------------------------------------------------
% FIGURE 11 
%-------------------------------------------------------------------------------

%
%-------------------------------------------------------------------------------
%
\subsection{The problem of the discrepant redshift}

Hanami et al. (\cite{hanami99}), fitting the redshift from the ASCA GIS and SIS
spectrum of \scnove, reported that the resulting value in not consistent with
the optical value, being $z_X<0.031$. They found that this discrepancy is 
essentially due to the fact that the centroid of the iron K$\alpha$ line
is more consistent with H-like  rather than He-like ionization. 
This result permitted  the authors to
suppose the presence also of gas at significantly higher temperature 
($kT \sim 15$ keV). 
\\
We fitted the 8 arcmin region by fixing all the parameters but the temperature, 
the redshift and the normalization, finding $z_{fit}=0.031$ with a
$90 \%$ interval covering the range [0.019-0.049]. The result does not
change by fixing the temperature.
However, we have to take into account a systematic shift of $-40$ eV in the 
channel-to-energy conversion (Fiore, private communication), resulting 
in a redshift shift of $+ 0.006$.
Note that, within the uncertaintes and this systematic shift, the 
$z_{fit}$ value is consistent with the mean redshift of both optical 
substructures (i.e. $0.044$ and $0.049$).
The result does not change by using data in a smaller region of $5$ arcmin
radius.
\\
As a further test, we fitted the spectrum with a bremsstrahlung$+$Gaussian
model, in order to determine the centroid of the line, finding 
$E=6.50$ keV with a $90 \%$ range of [6.38--6.63] keV: this result is at 
the border of consistency with the expected value of $6.37$ keV. Considering 
the correction of the $-40$ eV on the channel-to-energy systematic offset,
we found that the discrepancy is no more than  $2 \sigma$ of significance.   
As a check, we repeated this analysis also for \scsette: the resulting 
centroid is $E=6.37$ keV with a $90 \%$ range of [6.26--6.49] keV. 
\\
In conclusion, the evidence of presence of hotter gas in our data is weak,
even if in the same direction of that found by Hanami et al. (\cite{hanami99}),  
and a conclusive analysis will be possible only with the spectral capability 
of XMM-Newton.      

%
%-------------------------------------------------------------------------------
%
\section{LECS data}

Recently. Bonamente et al. (\cite{bonamente01}) claimed the detection in 
Beppo-SAX and ROSAT-PSPC data of a soft (at $E\sim 0.25$ keV) excess
in the clusters A3562, A3558, A3560 and A3571. The first two
clusters are part of the A3558 complex, while A3560 is $\sim 3.1$ \hmpc 
away from \scnove and $\sim 4.1$ \hmpc from \scsette and therefore could be
interacting with the structure. Moreover A3571 is part of another 
cluster complex of the Shapley Concentration.      
\\
In principle there are no indications that this soft excess is confined
in the single clusters: in fact it could be due to a diffuse, cold 
component (a filament) that we know to exist between A3558 and A3562
(Bardelli et al. \cite{bardelli96}, Kull \& B\"ohringer \cite{kull99}). 
In order to check the possible existence of a soft excess in our 
LECS data, we restricted the analysis in the $[0.2-0.4]$ keV band, considering
the central region of 8 arcmin radius. 
\\
We find that the excesses in this band, with respect to the counts expected 
from the best fit spectrum, are $0.5$ and $2.13$ $\sigma$ for 
\scsette and \scnove, respectively. Considering that the values of $\sigma$
are determined only by the counts statistics and not by the uncertaintes
on the fit and on the calibration, we conclude that no significant excess is 
present in our data.
Moreover, note that in Ettori et al. (\cite{ettori00}), where a Beppo-SAX 
observation of A3562 was analyzed, we have not found the excess reported by 
Bonamente et al. (\cite{bonamente01}). 

%
%-------------------------------------------------------------------------------
%
\section{PDS data}

There are evidences of the presence of hard X-ray excesses over a simple
bremsstrahlung model in some galaxy clusters which
present some degree of merging (see Fusco-Femiano \cite{fusco00}). This hard
energy tails have been detected for the first time by the PDS instrument 
on board of Beppo-SAX satellite.
\\
It is not clear which is the exact mechanism which accelerates the electrons 
and whether it is related with merging alone or if it requires some additional
parameters (as, for instance, a preexisting radiosource or a gas temperature 
hotter that 6-7 keV).  
\\ 
For this reason it is worthwhile to look if an hard tail is present in
our data. 
Following the data reduction cookbook, the main problems come from the
variable background of energetic particles, which has to be subtracted.
There are basically two methods to do this rejection, i.e. the  Fixed 
Rise Time threshold and the Variable Rise Time threshold. We checked
that the counts resulting after the application of the two methods are
consistent each other and give the same results. 
By imposing to the PDS data a {\it mekal} model with the same parameters fitted 
in the MECS alone, we find a significant excess in both \scsette and \scnove
observations, in particular between 15 and 30
keV. We found $6.59 \pm 2.51 \times 10^{-2}$ cts s$^{-1}$ against the 
$0.20\times10^{-2}$ cts s$^{-1}$ predicted by the {\it mekal} model for 
\scsette and  $5.34 \pm 2.07 \times 10^{-2}$ cts s$^{-1}$ 
(against $0.68\times10^{-3}$ cts s$^{-1}$) for \scnove.
The excess flux in the two observations is consistent within the errors
to be the same in the two exposures. 
The fit does not improve by adding a power law, which gives a formal slope 
$\sim 3$. 
\\
%
%-------------------------------------------------------------------------------
% FIGURE 12 
\begin{figure}
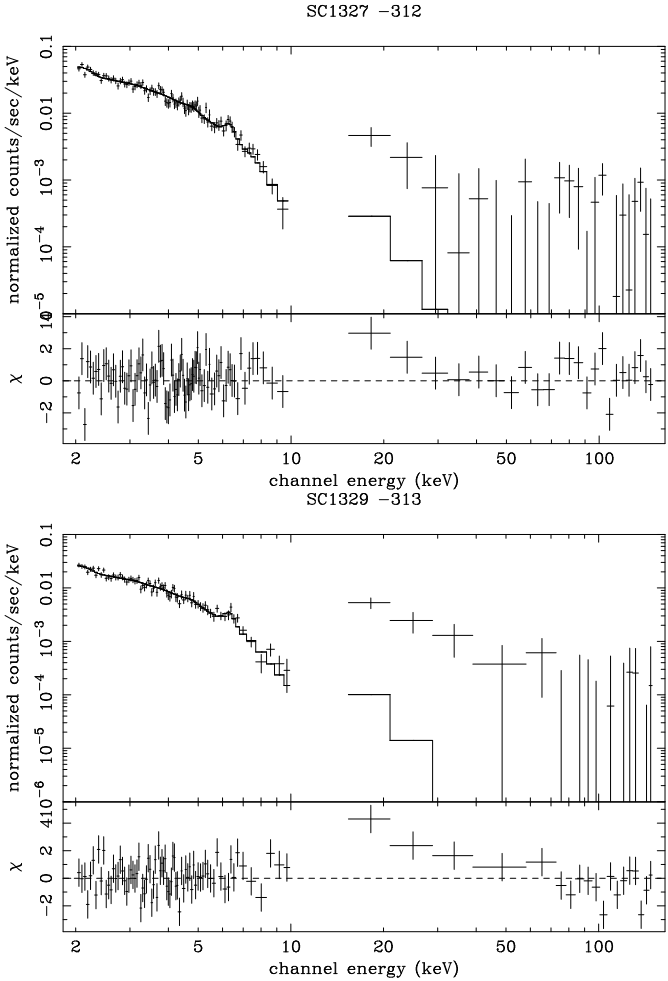

\centering
\includegraphics[angle=-90,width=\hsize]{paper_sc_fig12a.ps}
\includegraphics[angle=-90,width=\hsize]{paper_sc_fig12b.ps}
\caption{PDS+MECS spectrum of \scsette (upper panel) and \scnove 
(lower panel). }
\label{fig:PDS}
\end{figure}
%-------------------------------------------------------------------------------
The spectral shape of this excess
is inconsistent with the hypothesis of a cluster hard tail and indicates 
the presence of a contamination source.  
The problem in identifying the origin of this bump arises from the huge 
field of view of the PDS, i.e. $1^o.3$; our two exposures cover large
part of the A3558 complex. For this reason, added to the fact that
this region of the sky is poorly sampled by large X-ray surveys,
it is difficult to find the possible responsible for this excess. 
After having checked in the literature, we found
as a tentative candidate a QSO at $z=1.335$ (Drinkwater et al. 
\cite{drinkwater97}, Venturi et al. \cite{venturi00}), located 
at $\alpha_{2000}=13^{h} 30^{m} 19^{s}$
and $\delta_{2000}=-31^{o} 22' 59''$, i.e. North-East with respect to the
position of \scsette. This source was previously known as $PKS1327-311$
and presents a radio spectrum of steepness $\alpha_{13cm}^{22cm}=-1.13$ 
(with total flux of $S_{22cm}=421.7$ mJy, Venturi et al. \cite{venturi00}).
This source is in the field of view of the ROSAT-PSPC observation
of A3558 of Bardelli et al. \cite{bardelli96} with a $[0.5-2.0]$ flux of 
$16.2\times 10^{-14}$ erg cm$^{-2}$ s$^{-1}$.
\\
Finally, 
considering the response function of the PDS instrument, the count--rate
ratio between the \scsette and the \scnove observations is consistent
with having been originated by this source.
In fact, the observed ratio is $1.23\pm 0.67$, to be compared with 
the expected value of 1.32. 

%
%-------------------------------------------------------------------------------
%
\section{Discussion and conclusions}
 
The aim of this work has been the study of two blobs of hot gas (\scsette and
\scnove) positioned between A3558 and A3562, two ACO clusters in an oustanding 
phase of major merging. 
These clusters (together with A3556) form the A3558 complex, a structure 
with a linear dimension of $\sim 7$ \hmpc at a redshift of $\sim 0.05$.
The relative center-to-center distance between  A3562 and A3558 is
$\sim 3$ \hmpc.   
\\
The indication of ongoing merging comes both from the optical and from 
the X-ray band.
From the optical band, the envelope of galaxies sorrounding the clusters cores
(visible in Figure \ref{fig:iso}; see also figure 2 of Bardelli et al. 
\cite{bardelli98a} for the wedge diagram of the redshift sample) 
suggests that the galaxies from outer  
parts of A3562 and A3558 have been shared by the dynamical interaction. 
Also by studying the dynamics of the structure using redshift survey,
Bardelli et al. (\cite{bardelli00}) and Reisenegger et al. 
(\cite{reisenegger00}) concluded that 
the A3558 complex is at the late collapse phase. Moreover, the presence of 
a large number of subcondensations (Bardelli et al. \cite{bardelli98b}) 
reveals that the complex is far from the relaxation.  
\\
From the X-ray, both Bardelli et al. (\cite{bardelli96}) and Kull \& 
B\"ohringer (\cite{kull99}) detected diffuse emission extending along 
the whole A3558 complex, in the same way of the optical galaxies. 
Moreover, Markevitch \& Vikhlinin (\cite{markevitch97}) found regions of 
hotter gas with respect to the A3558 average temperature in the direction of 
\scsette, while Ettori et al. (\cite{ettori00}) detected 
diffuse gas outside A3562, on the opposite side with respect to the
A3558 direction.
This picture is consistent with the results of numerical simulations
of merging clusters.
\\
In order to study the role of \scsette and \scnove in this merging picture,
we analyzed two Beppo-SAX observations pointed on the centers of these two
poor clusters.
The MECS image of \scsette seems rather symmetric and centered on the 
dominant galaxy. On the contrary, the emission of \scnove has a ``comet-like" 
shape: we detected a significant excess of X-ray emission corresponding to
the position of the brightest galaxy (which appears to be double) and a tail
pointing eastward. 
In both poor clusters, it is visible the presence of diffuse emission
(see Figures \ref{fig:sc1327secmap} and \ref{fig:sc1329secmap})  
arising from the filament connecting the clusters of the A3558 complex.
Moreover, in \scsette and \scnove images a contribution from the X-ray 
emission of A3558 and A3562, respectively, can be detected.
These components are bright enough to allow a temperature estimate up to
a distance of 20 arcmin from the frame center. 
Finally, possible indications of disturbance come from the displacement 
between the positions of the X-ray peaks and the optical substructure 
centroids in both objects.  
\\
We estimated the ``global" (i.e. within 8 arcmin, corresponding to
$\sim 0.3$ \hmpc) temperatures and abundances for the two clusters 
using both MECS and LECS instruments. We find $4.11$ keV and $0.17$
for \scsette and $3.43$ keV and $0.21$ for \scnove.
These values are consistent with those expected from the velocity dispersions
of the corresponding optical substructures found in Bardelli et al. 
(\cite{bardelli98b}).  
\\
As a further step, we performed a spatially resolved spectroscopy with the
use of the MECS instrument. The temperature profiles are not very different 
from an isothermal distribution of the hot gas, a part from the fact that the
most internal bins in the two clusters are always colder than the others.
Then, we estimated the temperatures as a function of the distance from the 
center dividing the images in four sectors of 90$^{o}$. We rotated the grids 
in order to have a sector of \scsette pointing toward A3558 and   
another toward \scnove and a sector of \scnove  pointing toward A3562 and
another toward \scsette. 
\\
We found a good agreement between the temperatures derived in the overlapping
regions of the two pointings and in the regions at the edge with A3558 and 
A3562, where literature estimates are available. This confirms the 
goodness and reproducibility of the analysis of MECS images in the most
external regions, i.e. up to 20 arcmin. 
\\
We did not find evidence of the existence of a shock front 
inside the poor clusters, 
a part from a weak 
indication that in \scnove the temperatures in the bins between 2 and 8 arcmin
in the NW sector seem to be higher than the corresponding ones in the other 
sectors. 
There appears to be a temperature increase also in most external part of 
\scsette in the sector pointing toward A3558: however, this increase is
consistent with the hot region in A3558 found by Markevitch \& Vikhlinin
(\cite{markevitch97}). 
\\
An excess in the PDS energy range has been detected, but it resulted to be 
unrelated with the cluster emission and due to a contaminating source.
Moreover, analysing the LECS data, we did not find a significant soft excess
with respect to a single thermal bremsstrahalung.
\\ 
Finally, the presence of multiphase gas in \scnove, as claimed by Hanami et al. 
(\cite{hanami99}) on ASCA data, has been found only at the 2$\sigma$ confidence
level.   
\\     
In conclusion, we did not find evidence of the presence of shock-heated
gas in these two poor clusters. 
Only \scnove presents in its distorted isophotes sign of disturbance of the
hot gas: the gas profile of \scnove shows ``comet-like" shaped
isophotes, with the tail pointing toward A3562 and a compression 
toward \scsette.    
\\   
This is an indication that reinforces our work hypothesis: the A3558
complex is a remnant of a major merging seen after the first core-core
encounter. In the numerical simulations of mergers (see for example
Sarazin \cite{sarazin00}), the shock is created between two clusters before
the first collision. After the core-core encounter, the shock travels
toward the external regions of the clusters, where it will be dissipated.
The A3558 complex is therefore an important opportunity for studying evolved 
mergings and the new generation X-ray instruments (XMM-Newton and Chandra)
will be helpful in order to give a complete picture of this phenomenon. 

%
%-------------------------------------------------------------------------------
%
\begin{acknowledgements}

SB warmly thanks D.Dallacasa for his help with the deconvolution programs
and C.Vignali and S.Pellegrini for useful discussions.
We thank the referee for usefull comments. 
This research has made use of linearized event files produced
at the Beppo-SAX Science Data center.
This work has been partially supported by the Italian Space Agency grants
ASI-I-R-105-00 and ASI-I-R-037-01.

\end{acknowledgements}
%
%-------------------------------------------------------------------------------
%

%-------------------------------------------------------------------------------
%
\end{document}